\documentclass[sigconf]{acmart}

\AtBeginDocument{%
  }

\acmConference[Preprint]{Preprint}{}
\acmYear{}
\acmISBN{}
\acmDOI{}
\renewcommand\footnotetextcopyrightpermission[1]{} 

\usepackage{xcolor}

\definecolor{codegreen}{rgb}{0,0.6,0}
\definecolor{codegray}{rgb}{0.5,0.5,0.5}
\definecolor{codepurple}{rgb}{0.58,0,0.82}
\definecolor{backcolour}{rgb}{0.95,0.95,0.92}
\usepackage{braket}

\newtheorem{definition}{Definition}
\usepackage{balance}
\usepackage{listings}
\lstdefinestyle{mystyle}{
    commentstyle=\color{codegreen},
    keywordstyle=\color{magenta},
    numberstyle=\tiny\color{codegray},
    stringstyle=\color{codepurple},
    basicstyle=\ttfamily\footnotesize,
    breakatwhitespace=false,         
    breaklines=true,                 
    captionpos=b,                    
    keepspaces=true,                 
    numbers=left,                    
    numbersep=5pt,                  
    showspaces=false,                
    showstringspaces=false,
    showtabs=false,                  
    tabsize=2,
    frame=single,
}

\lstdefinestyle{ontop}{
  float=tp,
  floatplacement=tbp
}

\usepackage[utf8]{inputenc}
\usepackage{xspace}
\usepackage{algorithm}
\usepackage{stmaryrd}
\usepackage{colortbl}
\usepackage{cleveref}
\usepackage{tcolorbox}

\newcommand{\eg}{\emph{e.g.},\xspace}
\newcommand{\ie}{\emph{i.e.},\xspace}

\newcommand{\cf}{\emph{cf.}\xspace}
\newcommand{\aka}{\emph{a.k.a.},\xspace}

\newcommand{\coreutils}{\texttt{coreutils}\xspace}
\newcommand{\toybox}{\texttt{toybox}\xspace}
\usepackage{tikz}
\definecolor{lightGrey}{rgb}{0.9, 0.9, 0.9}
\newcommand{\Hilight}{\makebox[0pt][l]{\color{codegreen!10}\rule[-0.4em]{\linewidth}{1.5em}}}
\newcommand{\HilightB}{\makebox[0pt][l]{\color{yellow!50}\rule[-0.45em]{\linewidth}{1.5em}}}
\newcommand{\HilightC}{\makebox[0pt][l]{\color{codepurple!10}\rule[-0.45em]{\linewidth}{1.5em}}}
\newcommand{\HilightD}{\makebox[0pt][l]{\color{red!10}\rule[-0.45em]{\linewidth}{1.4em}}}
\newcommand{\HilightE}{\makebox[0pt][l]{\color{cyan!30}\rule[-0.5em]{\linewidth}{1.4em}}}

\begin{document}

\title{Small Yet Configurable: Unveiling Null Variability in Software}

\author{Xhevahire Tërnava}
\affiliation{%
  \institution{LTCI, Télécom Paris, Institut Polytechnique de Paris}
  \city{Palaiseau}
  \country{France}}
\email{xhevahire.ternava@telecom-paris.fr}

\author{Georges Aaron Randrianaina}
\affiliation{%
  \institution{Univ Rennes, CNRS, Inria, IRISA}
  \city{UMR 6074, F-35000 Rennes}
  \country{France}}
\email{georges-aaron.randrianaina@irisa.fr}

\author{Luc Lesoil}
\affiliation{%
  \institution{EssilorLuxottica}
  \city{Créteil, Île-de-France, France}
  \country{France}}
\email{luc.lesoil@irisa.fr}

\author{Mathieu Acher}
\affiliation{%
  \institution{Univ Rennes, CNRS, Inria, IRISA
  \\Institut Universitaire de France (IUF)
  }
  \city{UMR 6074, F-35000 Rennes}
  \country{France}}
\email{mathieu.acher@irisa.fr}

\renewcommand{\shortauthors}{Tërnava et al.}


\begin{abstract}
Many small-scale software systems, that is, with limited codebase or binary size, are widely used in everyday tasks, yet their configurability remains largely unexplored. At the same time, studies on modern software systems show a trend toward increasing configurability, alongside growing interest in building immutable, specialized, and reproducible software. In this paper, we present the first empirical study on the extent of configurability in small-scale software systems. By analyzing 108 programs from GNU \coreutils, we show that even small programs can exhibit significant compile-time and run-time variability, with up to 76 options per program. Then, there is a high correlation ($0.78$) between run-time variability and codebase size. Furthermore, an analysis of the 20 smallest programs across 85 releases reveals that variability tends to increase over time, primarily due to the added compile-time variability. This suggests that shifting options between run-time and compile-time, removing unnecessary run-time variability, or resolving compile-time variability early, can help reduce codebase complexity and size. We also introduce, for the first time, the concept of \textit{null-variable} software system, one with no configurability beyond mandatory features. Our findings show that high configurability is not exclusive to large-scale systems and that reducing unnecessary variability can lead to lightweight, smaller, and more maintainable software. We hope this effort contributes to designing new software by understanding how to balance its configurability with codebase size.
\end{abstract}





\maketitle

\section{Introduction}
\label{introduction}
Since the software crisis in 1968~\citep{mcilroy1968mass}, significant advances have been reached in the development of variability-intensive software systems. 
Nowadays, software systems are highly configurable and expose to users their abundant configuration options. These options make possible to customize a software system for usage in different contexts~\citep{capilla2013systems}, without the need to modify its source code. 
At the code level, most of the options are set at compile-time or run-time~\citep{capilla2013systems,bindingtime}, for example, as command-line parameters in C-based systems. Moreover, variability is often recognized as a strength of a software as users have the flexibility to adapt it to their needs.

However, software systems are becoming more and more configurable, such as Linux kernel with more than 20 thousand of options and a configuration space (\ie possible variants of Linux) of more than atoms in the universe~\citep{passos2016coevolution,martin2021transfer,10.1145/3729423,10.1145/3546932.3546997}. 
This ultra-large scale of variability is exceeding the human and even machine limits to deal with it, for example, to test and maintain it~\citep{10.1007/s10664-018-9635-4,10.1145/2642937.2642990,yin2011empirical}.
The ultimate and reasonable goal of researchers and developers today is how to address,
manage, evolve, and reduce this large and increasing software variability. 
To our knowledge, the extent of variability and configurability in small-scale software is seldom analyzed, and the notion of a completely invariant system remains undefined.

Actually, motivated by this gap, we asked several questions.
What is the extent of variability in small-scale software systems, compared to the extensive configurability of large-scale systems? Is configurability also an important characteristic of small-scale software? Do software systems exist that exhibit little to no noticeable variability?
There are studies that have used small-scale software, simplified versions of larger systems, or excerpts from their feature models, for various purposes, such as evaluating variability implementation approaches~\cite{apel2016feature,10.1007/3-540-44800-4_2} or debloating software systems~\cite{brown2024broad}. 
But, in our knowledge, there is no work that explores such questions. 
In addition, the field of software engineering has traditionally focused on configurability and adaptability, leaving systems with no variability largely unexplored. While terms like "static", "immutable", or "constant" software are occasionally used\footnote{https://sonalake.com/latest/the-future-is-minimal-and-immutable-a-new-generation-of-operating-systems/}, there is no formal concept to describe systems designed to lack configurability, representing the antithesis of configurable systems.

Herein, we empirically investigate the extent of variability in small-scale software by analyzing 108 programs from \coreutils\footnote{GNU core utilities: \url{https://www.gnu.org/software/coreutils/}} and 85 releases across 20 of its programs spanning the years 2003 to 2022, including programs such as \texttt{ls}, \texttt{cat}, and \texttt{mkdir}. Most of these programs have implementation histories dating back to 1979~\citep{inflation2}.
Our findings show that even small-scale software programs exhibit significant variability, with up to 76 configuration options and an average configuration space of $2^{15}$. Moreover, we observed that run-time and compile-time variability have distinc roles in the dynamics of software systems. 
While runtime variability impacts the size of a program's codebase at a specific point in time, compile-time variability appears to drive the growth of the codebase over time.
Additionally, we identify two strategies for reducing codebase size, both centered on removing unnecessary runtime variability.

The main contributions of this paper are as follows:
\begin{itemize}
    \item We present the first empirical study on the extent of variability in small-scale software, unlike previous studies that focus on large, complex configurable software systems or explore small-scale software for other purposes.
    \item To the best of our knowledge, we are the first to explicitly introduce and define the concept of \emph{null-variable} software.
    \item All the data and code to replicate our study are available at \textcolor{blue}{\url{https://github.com/ternava/zero-variability}}. %
\end{itemize}

The remainder of this paper is structured as follows. We first provide the background (\Cref{background}) and  motivation (\Cref{motivation}). Then, we present our empirical study (\Cref{studydesign}) with four research questions, examining the size and variability of over 108 programs. We report on the variability, correlation with system size, evolution, and reduction in~\Cref{results}, and discuss the results in~\Cref{discussion}. In~\Cref{definitionnull} we define \textit{null software variability}, threats to validity in~\Cref{threats}, related work in~\Cref{related}, and conclude in~\Cref{conclusion}.

\section{Background}
\label{background}

\paragraph{Variability-intensive software systems.}
Software systems that are built nowadays require and exhibit variability, commonly understood as \emph{the ability of a software system or artifact to be efficiently extended, changed, customized, or configured for use in a particular context}~\cite{capilla2013systems,svahnberg2005taxonomy}.
Software variability is often studied in the context of software reuse~\cite{mcilroy1968mass}, software factories, software families, software product lines (\eg~\cite{kang1990feature,capilla2013systems,apel2016feature}), and configurable systems (\eg~\cite{gazzillo2022bringing,ternava2022scratching}). The last two abilities,~\ie~\textit{customizability} and \textit{configurability}, are typically described in terms of \textit{features} or \textit{variation points with variants} incorporated into their respective feature or decision models~\cite{kang1990feature,czarnecki2012cool,ternava2022identification}.
Similarly, \textit{configuration options} are used in configurable software systems, turning them into families of closely related software systems~\cite{gazzillo2022bringing}.
Moreover, all software artifacts, such as architecture, design, and code, have the potential to vary or to be reused, whether through ad-hoc or systematic reuse. 
A recent study shows that code artifacts are the most frequently reused ones~\cite{capilla2019opportunities}. Therefore, in this work, we focus on exploring variability of small-scale configurable software at the code level.

\paragraph{Binding time.}
Although a software system can change over time (\aka~variability in time), variability-intensive systems excel in their ability to do and be a dozen different things at a certain point in time (\aka~variability in space)~\cite{clements2019managing}. 
This variability is initially introduced, for example, in the form of options, and then resolved (\ie~concretized)~\cite{svahnberg2005taxonomy}, either immediately or later in the development cycle. 
The point in time at which some variability is resolved is known as the binding time and can range from the requirements phase to runtime~\cite{capilla2013systems,krisper2016describing,bindingtime}. At the code level, all binding times are often boiled down to the \textit{programming time}, \textit{compile time}, and \textit{runtime}.  In this work, these are also the binding times of variability in small-scale software systems that we focus on examining.

\paragraph{Implementation: examples of variability types and systems analyzed}
\Cref{lsprogram} shows an excerpt of the implementation of three well-known C-based command-line software programs: \texttt{ls}, which lists directory contents; \texttt{dir}, which briefly lists directory contents; and \texttt{vdir}, which verbosely lists directory contents\footnote{Manual: \url{https://www.gnu.org/software/coreutils/manual/coreutils.html}}. Because of their commonality in the behavior, they are implemented together. Specifically, the compile-time variability in lines 2-4, realized by preprocessor directives, makes it possible to build the three programs with their own executables from the same C files.
Whereas, their run-time variability in lines 6-21 is achieved through the \texttt{getopt.h} library\footnote{\url{https://www.gnu.org/software/libc/manual/html_node/Getopt.html}}, which automates the parsing of Unix command-line options.
The program's usage, as shown in~\Cref{lscoreutils} (line~2), and all its available run-time options can be printed in the standard output (lines 3-6) by calling the program with the \mbox{-}\mbox{-}\texttt{help} option (line~1).
\begin{lstlisting}[style=ontop,escapechar=\%,caption={An excerpt of \texttt{ls.c} of \texttt{ls}, \texttt{dir}, and \texttt{vdir} programs}, label=lsprogram, language=C]
#include <getopt.h>
#define PROGRAM_NAME (ls_mode == LS_LS ? "ls" \
                      : (ls_mode == LS_MULTI_COL \
                         ? "dir" : "vdir"))
int main (int argc, char **argv) {                         
while (true) {
      int c = getopt_long (argc, argv,
        "abcdfghiklmnopqrstuvw:xABCDFGHI:LNQRST:UXZ1",
        long_options, &oi);
      if (c == -1)
        break;
      switch (c) {
        case 'd':
          immediate_dirs = true;
          break;
        case 'l':
          format_opt = long_format;
          break;
          /* The rest of code and options are omitted */
        }
    }
}
\end{lstlisting}
\begin{lstlisting}[style=ontop,escapechar=\%, caption={Usage and run-time options of the \texttt{ls} program}, label=lscoreutils]
%\HilightE%[coreutils9.1]$ ./src/ls --help
>> Usage: ./src/ls [OPTION]... [FILE]...
>> -d, --directory   list directories themselves, 
>>                   not their contents
>> -l                use a long listing format
%\Hilight%The rest of 58 options are omitted
\end{lstlisting}

Although many approaches for addressing variability exist, such as feature-oriented or software design patterns~\cite{svahnberg2005taxonomy,apel2016feature,ternava2017diversity}, this study focuses on the type of variability found in C-based configurable systems, specifically analysing their compile-time and run-time variability.
Moreover, we refer to such systems, characterized by their small executable size, as small-scale systems in this study.
\begin{lstlisting}[escapechar=\%, caption={Usage of a compile-time option to build a program}, label=greputility]
%\HilightE%[grep-3.10]$ ./configure %\textcolor{blue}{\mbox{-}\mbox{-}disable\mbox{-}largefile}%
[grep-3.10]$ make && make install
\end{lstlisting}
\begin{lstlisting}[escapechar=\%, caption={Usage of a run-time option and input in a program}, label=greputility2]
%\HilightE%[grep-3.10]$ ./src/grep %\textcolor{blue}{\mbox{-}\mbox{-}ignore\mbox{-}case}% %\textcolor{codegreen}{"novel" books.txt}%
\end{lstlisting}
\paragraph{Configuration: illustration of compile-time and run-time variability}
Often, to customize a C-based software at compile time, users simply use the \texttt{./configure} command~\cite{autoconf}. For example, \texttt{grep}\footnote{\texttt{grep}: \url{https://www.gnu.org/software/grep/}} is a program that is used to search for matching patterns in a file. It has dozens of configuration options that can be used to customize its behavior before installation. For instance, in case that the support for large files is not required, then it is possible to deactivate it by using the \texttt{--disable-largefile} option during the \texttt{grep}'s build, as in~\Cref{greputility} (line 1). After its build, numerous run-time options become available. For example, to search for the word ``novel" in a case-insensitive manner in a file, a user can use the command shown in~\Cref{greputility2}. In this case, \texttt{./src/grep} is the program, \mbox{-}\mbox{-}\texttt{ignore}\mbox{-}\texttt{case} is the run-time option that affects program's behavior during a particular execution, and \texttt{``novel" books.txt} is the program's input. This study focuses on this kind of compile-time and run-time variability often found in C-based software systems.

\section{Motivation}
\label{motivation}
Studies show that software systems are becoming increasingly configurable to meet user needs and cope with the heterogeneity of execution platforms. For example, the Linux kernel, with over 20\,000 features and 26 million lines of code, offers a configuration space larger than the number of atoms in the universe~\cite{10.1145/3729423,10.1145/3546932.3546997}, a number projected to double within two decades~\cite{10.1145/3546932.3546997}. This immense configuration space poses significant challenges, such as testing a representative subset of configurations~\cite{sampling, 10.1007/s10664-018-9635-4}, identifying bugs caused by feature interactions~\cite{10.1145/2642937.2642990, 10.1145/3149119}, and performing variability-aware testing~\cite{10.1145/3280986, 10.1145/2568225.2568300}. Even basic tasks, such as counting all possible configurations, become very complex~\cite{10.1145/3546932.3546997}. However, there is little empirical evidence on the scale of variability in other typical real-world software systems~\cite{10.1145/3307630.3342400,10.1145/1806799.1806819} or how it compares to larger ones like Linux kernel. On the other hand, many small-scale software systems, characterized by their limited codebase or binary size, such as GNU \coreutils, are widely used in everyday tasks. However, to the best of our knowledge, their configurability remains largely unexplored. 
Therefore, we question whether variability is also an important characteristic of small-scale systems, as in large-scale ones, or if their limited configuration spaces would make challenges like counting and testing feature combinations less relevant.

In fact, both real-world and artificially developed small-scale codebases, or snippets of configuration spaces of larger software systems, are often used by researchers, primarily for evaluation purposes, such as assessing new variability implementation techniques~\cite{10.1007/3-540-44800-4_2,10.1007/11554844_3,apel2016feature}, variability reverse engineering~\cite{10.1145/3377024.3377037}, or debloating approaches~\cite{brown2024broad}. The configuration space of these systems is often perceived as less complex to understand, although there is a lack of empirical evidence that supports this claim. We found no studies that specifically explore the configurability of small-scale systems, both in time and space, and its relationship to codebase size. Exploring the variability and configurability of small-scale systems can provide valuable, empirically grounded insights that we believe that could be effectively applied to larger-scale software systems. 

Today, from operating systems to programs like \coreutils, there is a clear trend towards specializing, debloating, and minimizing software systems to meet real-world demands for greater efficiency~\cite{10.1145/3689937.3695792}. Additionally, the push for more reproducible software emphasizes the need for minimal and immutable designs\footnote{https://sonalake.com/latest/the-future-is-minimal-and-immutable-a-new-generation-of-operating-systems/}. Despite these efforts, we notice a gap in understanding what defines or characterizes a \emph{null-variable} software system.
Moreover, we believe that understanding the variability of smaller systems can help address challenges in both small- and large-scale contexts.

\section{Empirical Study}
\label{studydesign}
In this section, we present the research questions and the $108$ small-scale software systems used as subjects to address them.

\subsection{Research Questions}
\label{researchquestions}
To examine the dynamics of variability in small-scale software systems from the user’s perspective, we asked:
\begin{description}
    \item[\textbf{$RQ_1:$}] \label{RQ1}
    \textbf{What is the extent of variability in small-scale software?}
    The aim of this research question is to analyze and quantify the variability in small to tiny software systems. Specifically, we measure variability by enumerating compile-time and run-time configuration options across 108 programs in the recent version of GNU \coreutils.
    \item[\textbf{$RQ_2:$}] \label{RQ2}
    \textbf{Is there a relationship between software size and variability?}
    Building on the results from the previous question, this research question investigates the relationship between software variability and size. To achieve this, we analyzed the binary size and lines of code alongside the amount of variability, calculating the Pearson and Spearman correlations for 108 programs in GNU \coreutils to explore the nature of this relationship in small-scale software systems.
    \item[\textbf{$RQ_3:$}] \label{RQ3}
    \textbf{How has variability in small-scale software changed over time?}
    The goal of this research question is to determine whether the earliest versions of small-scale software exhibited lower variability. To investigate this, we selected 20 programs from GNU \coreutils that currently have the lowest variability and analyzed how their compile-time and run-time configuration options evolved over time. 
    \item[\textbf{$RQ_4:$}] \label{RQ4}
    \textbf{What approaches do developers use to reduce or remove variability in small-scale software?}
    Systems like \toybox were created to provide lightweight alternatives to GNU programs, minimizing size and complexity. Our intent is to explore the strategies developers employ to minimize variability.
    To address this question, we compare alternative implementations of 3 programs in \coreutils and \toybox.
    Through this comparison, we aim to uncover some effective approaches for reducing variability in small-scale software.
\end{description}

\subsection{Subject Systems}
\label{subjects}
To answer the research questions, we analysed the \coreutils of GNU operating system.
They are the basic file, shell, and text manipulation utilities that are expected to exist on every operating system\footnote{GNU core utilities: \url{https://www.gnu.org/software/coreutils/}}. 
Specifically, our software subjects are 108 utilities in the current version 9.1 of \texttt{coreutils}\footnote{The source of analysed release: \url{https://mirror.cyberbits.eu/gnu/coreutils/}}, to which we refer also as programs.
Examples of such programs are 
\texttt{cat}, which makes possible to concatenates and writes files;
\texttt{echo}, which prints a line of text;
\texttt{ls}, which lists directory contents;
and \texttt{mkdir}, which makes directories.
Their historical relevance makes \coreutils a cornerstone of GNU/Linux systems, representative of traditional UNIX-like utilities.
The main reasons why we chose to analyze them are the fact that they are quite well-known, used in daily bases (\eg \texttt{ls}, \texttt{rm}), often used to evaluate different approaches in software engineering, such as~debloating techniques~\cite{brown2024broad}, and most importantly their relatively small codebase size (\eg compared to web applications\footnote{The website obesity crisis: \url{https://idlewords.com/talks/website_obesity.htm}}), in terms of lines of code (LoC) or binary size, makes these system suitable to the purpose of our study.

\begin{figure}[t]
    \centering
    \includegraphics[width=\columnwidth]{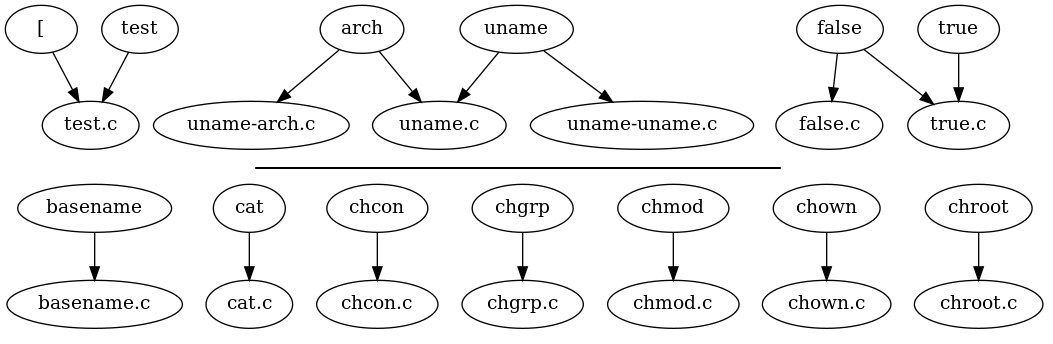}
    \caption{Dependency graph between programs and \texttt{.c} files}
    \label{fig:dep_graph}
\end{figure}

\subsection{Data Extraction and Processing Protocol}
\label{protocol}
\paragraph{The build of programs}
To analyse the programs of \coreutils, we had first to understand the way how they are implemented and can be build. In the considered release, the project of \coreutils has 763,637 LoC\footnote{We used \texttt{cloc} to count them: \url{https://github.com/AlDanial/cloc}}. One characteristic of \coreutils is that each program can be build as a separate binary (\ie executable), which is the default way, or the whole \coreutils can be build as a single binary containing the chosen programs. In this work, we built all the default programs offered by the project separately, meaning each has its own distinct binary. Note that two programs (\texttt{arch} and \texttt{hostname}) were not built by default, but we added them to the configuration. Despite that, we kept unchanged the default compile-time options during the build. The entire \coreutils resembles an SPL with the basic programs for an operating system. Thus, each of these programs can be seen as a product variant of \coreutils. 

\begin{figure*}[t]
    \centering
    \includegraphics[width=\textwidth]{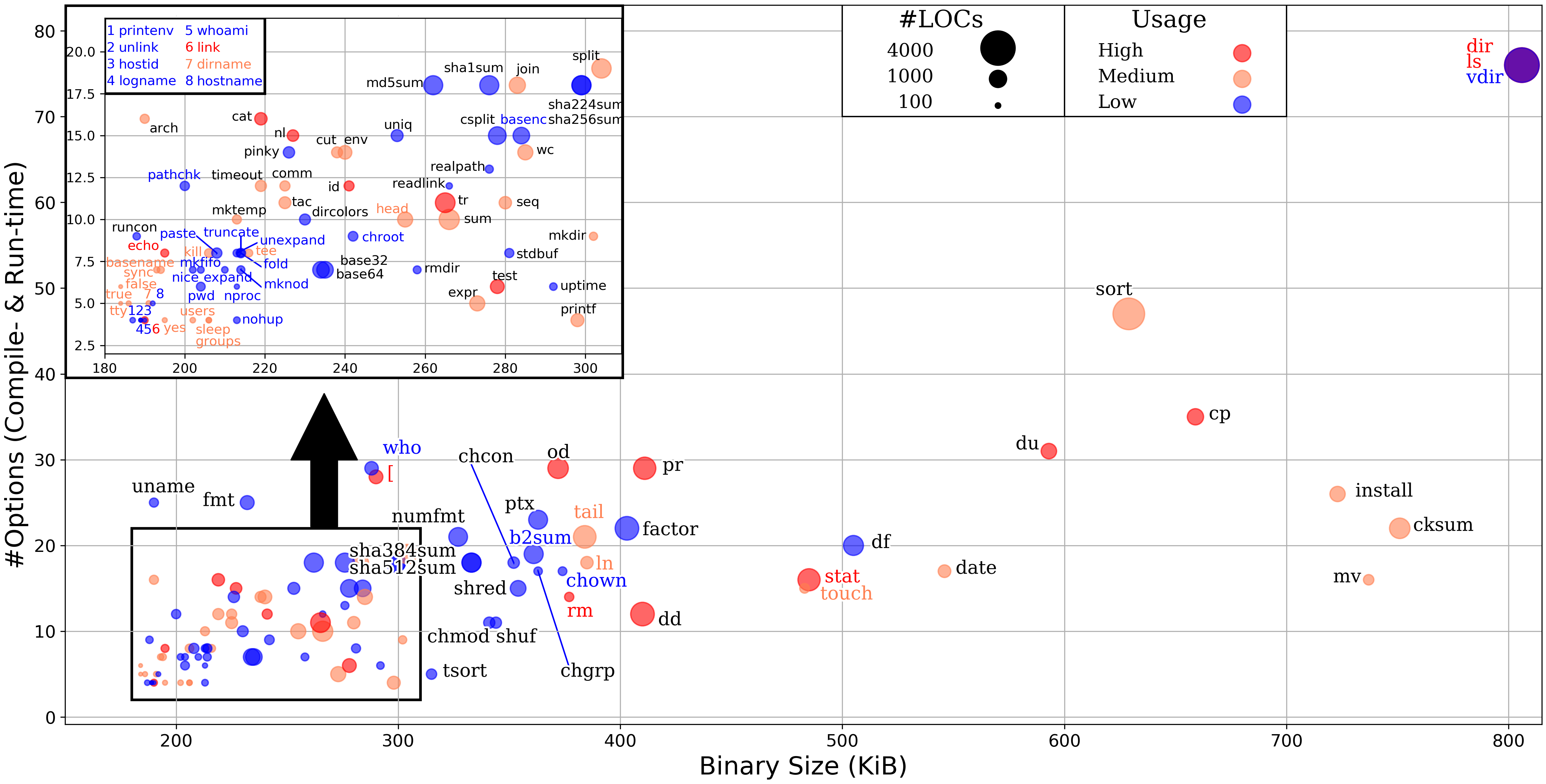}
    \caption{The variability, binary size, LoC and usage of 108 programs, as subjects, from the GNU \coreutils, version 9.1}
    \label{fig:bubbles}
\end{figure*}

\paragraph{Measuring the size of programs.}
Because of the product line nature of \coreutils, finding all used lines of code (LoC) to implement each program was challenging. Therefore, as the LoC of a given program we considered only the source code that explicitly contributes to the implementation of its functionalities. Specifically, for each program, we identify its main \texttt{.c} file, which is typically named after the program. Additionally, we include the LoC of files that implement features specific to the program. This analysis gives us a dependency graph as depicted in~\Cref{fig:dep_graph}. It shows that each program has often only one associated \texttt{.c} file (bottom part), but there are also programs with more complex dependencies (top part). For example, for the \texttt{cat} program we counted LoC that are in its corresponding \texttt{cat.c} file, for \texttt{arch} we counted LoC in \texttt{uname-arch.c} and \texttt{uname.c}, and so on. We did not count the LoC used to implement some functionalities which are imported as external files into the 
respective 
files of a program. Moreover, to increase the precision of our experiments and crosscheck the results, we also measured each program's executable size (in bytes). 

\paragraph{Quantifying variability in programs}
To determine the \textit{run-time variability} of a program, we first built the program and enumerated the number of individual options, being them long or short, shown by its \mbox{-}\mbox{-}\texttt{help} command. Duplicate and alias options (\eg \mbox{-}\texttt{a} and \mbox{-}\mbox{-}\texttt{archive} in the \texttt{cp} program, which provide the same functionality to the users) were counted as a single option. Similarly, variants of an option (\eg~the \texttt{basename} program's \mbox{-}\mbox{-}\texttt{suffix=SUFFIX} option) were counted as a single option because their possible values, such as different \texttt{SUFFIX}-es, are often unlimited and rarely a finite set.  Additionally, we disregarded logical relationships between options, such as alternative or optional dependencies, and all other kinds of dependencies between options. In other words, we counted the number of distinct available run-time options in a program, not its configuration space (\ie the possible combinations of options, or system variants). 
As for \textit{compile-time variability}, we used the graph in~\Cref{fig:dep_graph} to parse the source code of each program and automatically searched for preprocessor directives with Tree-sitter~\cite{syntax_tree}\footnote{Tree-sitter parser:  \url{https://tree-sitter.github.io/}}. We extracted each preprocessor directive and counted the distinct ones to determine the compile-time options available in each program.

\paragraph{Categorizing programs based on their usage}
\label{categorization}
To give some perspective on the most and least used programs in \coreutils, we extracted all the bash files used in the Linux kernel\footnote{\url{https://github.com/torvalds/linux/commit/a185a0995518a3355c8623c95c36aaaae489de10}} and counted the occurrences of each of the 108 subject programs within these files. We then classified the usage into three categories: \textit{High} for more than 1\,000 occurrences, \textit{Medium} for 50 to 1\,000 occurrences, and \textit{Low} for fewer than 50 occurrences of each program. 

\section{Results}
\label{results}
This section presents results, insights, and discusses the findings related to our four research questions.

\subsection{Extent of Variability in Small Software}
\label{answerRQ1}
To answer the first two research questions (\hyperref[RQ1]{$RQ_1$} and \hyperref[RQ2]{$RQ_2$}), 
we first measured the size of the 108 programs using two metrics: the number of lines of code (LoC) and their binary size in kibibytes (KiB). We then quantified their variability by enumerating the number of compile-time and run-time options, a common and immediate indicator of variability in C-based systems~\cite{el2019metrics} like \coreutils. 

In~\Cref{fig:bubbles}, the size of each bubble represents the obtained size of the corresponding program, measured in lines of code (LoC). It resulted that three programs, namely \texttt{dir}, \texttt{ls}, and \texttt{vdir}, have the highest number of LoC (4\,095~LoC), whereas the smallest ones are \texttt{hostid} and \texttt{true} with only 50~LoC. Thus, on average, a program has 712~LoC. As for the binary size, all programs fall within the range of $185$~[KiB] (for \texttt{true} and \texttt{false}) to $806$~[KiB] (for \texttt{dir}). This is shown on the horizontal axis in~\Cref{fig:bubbles}. On average, the size of the executable of a program is $308$~[KiB]. To put this in perspective, these sizes are small compared to many modern applications, \eg web applications, which often exceed several mebibytes (MiB)\footnote{The website obesity crisis: \url{https://idlewords.com/talks/website_obesity.htm}}. Hence, our subjects are indeed lightweight or small-scale software.

The vertical axis in~\Cref{fig:bubbles} also shows the 
amount of variability observed in the 108 analysed programs in \coreutils. 
To facilitate comparison in the chart, we categorized all programs based on their usage in the bash files of Linux kernel, as explained in~\Cref{categorization}. It resulted that approximately $16\%$ are highly used (\tikz\draw[red,fill=red] (0,0) circle (.5ex);, in red), $33\%$ are moderately used (\tikz\draw[orange,fill=orange] (0,0) circle (.5ex);, in orange), and $51\%$ are less used (\tikz\draw[blue,fill=blue] (0,0) circle (.5ex);, in blue). 

\begin{table}[t]
\caption{Correlation between variability, LoC, and binary size}
\label{tab:corrs}
\setlength{\tabcolsep}{2.5pt}
{\renewcommand{\arraystretch}{0.75}
\begin{tabular}{@{}lccr}
\toprule
Correlation & Pearson & Spearman & Resolution
\\ \midrule
Binary size $\times$ Run-time & 0,78 & 0,70 & Strong [0,6$\to$0,8] \\
LoC $\times$ Run-time & 0,78 & 0,62 & Strong [0,6$\to$0,8] \\
Binary Size $\times$ LoC & 0,72 & 0,70 & Strong [0,6$\to$0,8] \\ 
Binary Size $\times$ Compile-time & 0,38 & 0,38 & Low [0,2$\to$0,4] \\ 
LoC $\times$ Compile-time & 0,64 & 0,61 & Strong [0,6$\to$0,8] \\ 
Run-time $\times$ Compile-time & 0,35 & 0,23 & Low [0,2$\to$0,4] \\ 
\bottomrule
\end{tabular}
}
\end{table}

In terms of variability, all programs were found to have a range of 4 to 76 compile-time and run-time options.
Indeed, we found out that 3 programs, namely \texttt{dir}, \texttt{ls}, and \texttt{vdir}, have the largest amount of variability, with 76 compile-time and run-time options each. Whereas, 12 programs have only 4 compile-time and run-time options, among which are \texttt{sleep}, \texttt{true}, and \texttt{link}.
Although less significant, each program, on average, has 15 options.
If each of these options is Boolean (as is often the case), the configuration space for a program with 15 options would be $2^{15}$, or 32\,768 possible configurations, assuming that there are no dependencies between the options.
Thus, under these assumptions, even small programs offer substantial configurability, ranging from $2^{4}$ to $2^{76}$ possible configurations. This means that although small in binary size and LoC, still these systems can pose considerable challenges in testing and maintenance due to their large configuration spaces.
When compared to larger systems, this variability becomes even more striking. Studies on 26 larger software systems, such as x264, FreeBSD, and Ubuntu, report feature counts ranging from 9 to 31\,012, with an average of 2\,370 features per system~\cite{10.1145/3307630.3342400}. While these studies do not claim to provide exact counts of features~\cite{10.1145/3729423}, they reveal the immense configurability in real-world software systems. Our findings highlight that the configurability in smaller software like \coreutils is lower but significant and represents an important aspect that should not be overlooked when building, using, or maintaining them.

\begin{tcolorbox}[boxsep=-2pt]
\textbf{\hyperref[RQ1]{$\boldsymbol{RQ_{1}}$} insights:}  
Small to tiny software programs (up to $4\,095$ LoC and $806$ KiB in executable size) exhibit significant variability, with configuration options ranging from 4 to 76. Variability remains a key characteristic of these programs, enabling substantial configurability despite their modest size.
\end{tcolorbox}

\begin{figure*}[t]
    \centering
    \includegraphics[width=\linewidth]{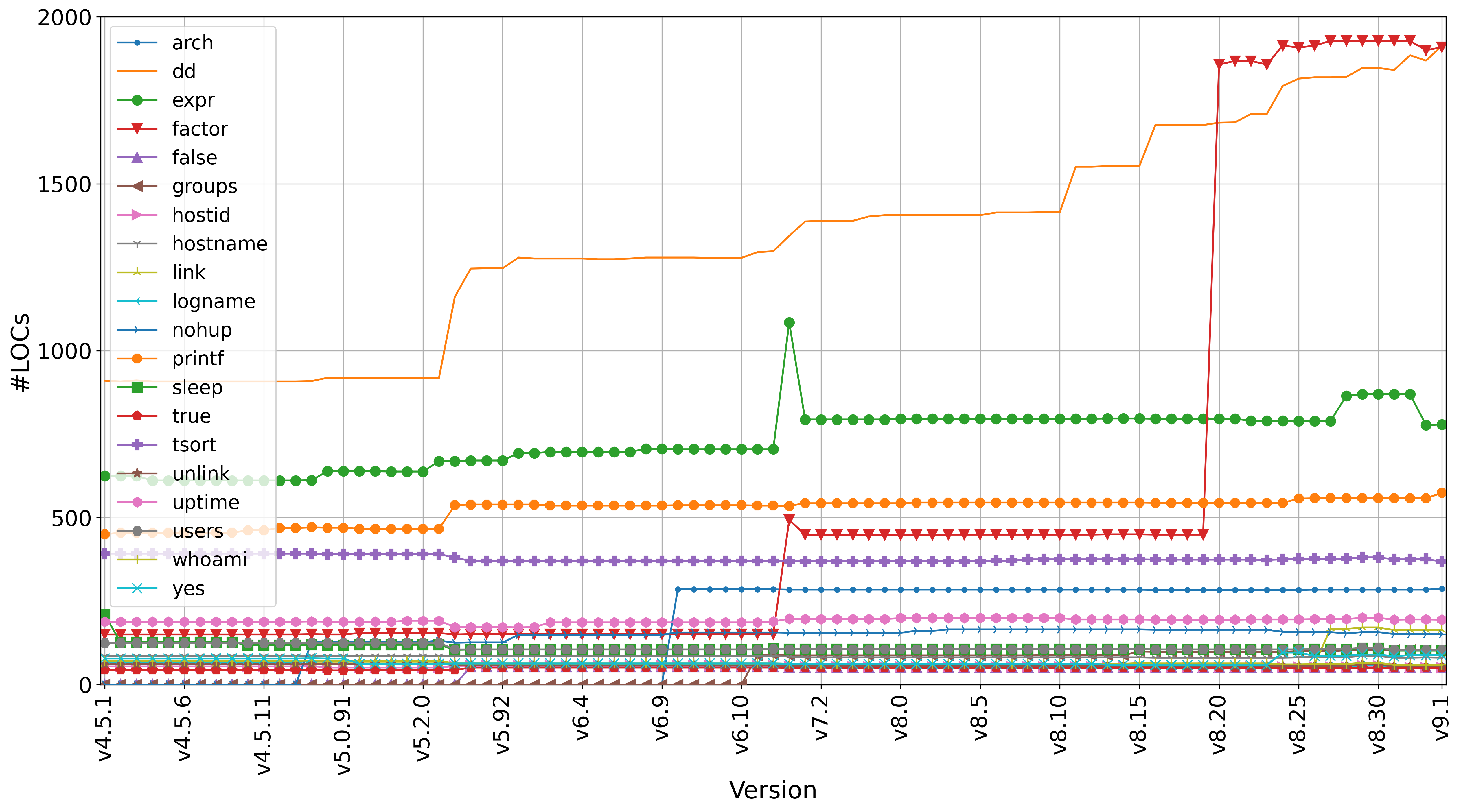}
    \caption{Changes in LoC over time for the 20 smallest programs, from 2003 (v 4.5.1) to 2022 (v 9.1)}
    \label{fig:evolution}
\end{figure*}

\subsection{Software Size and Variability Relationship}
Looking in~\Cref{fig:bubbles}, the majority of programs are concentrated in one area. Specifically, most of them have fewer than 20 run-time options ($91\%$), fewer than 10 compile-time options ($91\%$), an executable size of less than 500 [KiB] ($90\%$), and fewer than 1,500~LoC ($91\%$, as shown by the small bubbles). To verify if there is a relationship between these three measurements, we calculated their linear correlation using the Pearson and Spearman coefficients interpreted with the Evans rule~\cite{spearman}. 
The obtained results in~\Cref{tab:corrs} show that a high and positive ($\approx0.72$, that is, the average of the first two rows) correlation exists between the size (\ie binary size and $\#$LoC) and number of run-time options in programs.
Whereas, a lower and positive ($\approx0,50$, that is, the average of the fourth and fifth rows) correlation is between size and compile-time options. 
The highest correlation of $0.78$ is between binary size and run-time options. This suggests that a high and positive relationship exists between them, indicating that larger programs tend to have more run-time options. However, while the correlation is high, it is not perfect. 
This implies that the amount of variability in programs has an impact on their codebase size, or vice versa, alongside other factors. These results show how design choices and system architecture affect the balance between run-time and compile-time variability. This balance, in turn, affects the software's size and configurability.
\begin{tcolorbox}[boxsep=-2pt]
\textbf{\hyperref[RQ2]{$\boldsymbol{RQ_{2}}$} insights:}
The results provide an interesting perspective on balancing software size and configurability: while run-time options correlate strongly with size (0.78), compile-time options show a weaker correlation (0.50). This suggests an opportunity to strategically decide when to implement options at compile-time or run-time. Understanding this balance is key to optimizing configurability in systems of all sizes.
\end{tcolorbox}

\subsection{Variability Over Time in Small Software}
\label{variabilityovertime}
As results in~\hyperref[RQ2]{$RQ_2$} show, none of the programs are without variability. 
Then, there is a high correlation between their size and run-time variability (on average $0.78$ for Pearson and $0.66$ for Spearman). 
Therefore, for our third research question (~\hyperref[RQ3]{$RQ_3$}), we sought to determine whether earlier versions of these programs exhibited more or fewer variability. This is particularly relevant because it is possible that more lines of code were added to the programs over time, which may or may not be linked to new added options in the programs. To find out, we analysed the evolution of 20 programs (from 22 in total\footnote{Two of them have operands, another form of variability, besides options.}) that are with 2 run-time options. These programs have the lowest amount of variability in the current \coreutils. As such, we thought that they were ideal for closely examining, and even manually if needed, the relationship between variability and program size.
These programs are listed in the legend of~\Cref{fig:evolution}. Specifically, we analysed their run-time variability and size (LoC) in all 85 releases of \coreutils, from 2003 (v 4.5.1) to 2022 (v 9.1).

\begin{lstlisting}[caption={The three variants of the \texttt{true} program}, label=truerun, language=C]
$ ./src/true /*returns true as exit status*/
$ ./src/true --help /*displays its help and exits*/
$ ./src/true --version /*outputs its version*/
\end{lstlisting}
After a thorough examination of the variability in these 20 programs, we observed that all of them share the same 2 options in the latest release, namely  \mbox{-}\mbox{-}\texttt{help} and \mbox{-}\mbox{-}\texttt{version}. These two options seem that do not alter the functionality of these programs and are not particular to them, as pretty much all other programs also have them. We thus raised the question of whether they should be regarded as program variability at all. For instance,~\Cref{truerun} shows that both options make 3 possible ways to run \texttt{true}, and thus the 3 different outputs. Since they introduce a form of variability in the output, we considered them part of the program's variability.

\Cref{fig:evolution} shows the evolution of the number of lines of code ($\#$~LoC) for the 20 smallest programs in \coreutils. It can be noticed that their size ($\#$~LoC) decreased on average up to the version~5.0 in 2003. From this version onward, up to the current version~9.1, their size has increased by up to $99.89\%$ compared to the first version. 

Regarding the evolution of their variability, we noticed the following.
\begin{itemize}
    \item Programs \texttt{arch}, \texttt{false}, \texttt{groups}, and \texttt{nohup} were introduced in latter versions. For instance, the command \texttt{false} existed from the beginning as part of the built sources, but as a program in its own source file and executable, it was introduced only in version 5.90.
    \item Programs \texttt{expr} and \texttt{factor} have an increase in variability only in version 7.0, with 2 and 1 run-time options added, respectively. Then, their variability was decreased (\ie removed) by the same amount. 
    \item For the 13 other programs, their run-time variability remained the same across all releases (\ie with 2 options).
\end{itemize}

\begin{lstlisting}[escapechar=\%,caption={An excerpt of \texttt{true}, comparing its v 9.1 and v 5.0}, label=truefalse, language=C]
#include <stdio.h>
#include "system.h" /* ...*/

/* some of the code is omitted */
%\HilightB%#if EXIT_STATUS == EXIT_SUCCESS   /* true 9.1 */
# define PROGRAM_NAME "true"      /* true 5.0 */
%\HilightB%#else                             /* true 9.1 */
%\HilightB%# define PROGRAM_NAME "false"     /* true 9.1 */
%\HilightB%#endif                            /* true 9.1 */

int main (int argc, char **argv)
{ /* ...*/
  if (argc == 2)
    { /* ...*/
      if (STREQ (argv[1], "--help"))
        usage (EXIT_STATUS);

      if (STREQ (argv[1], "--version"))
        version_etc (stdout, PROGRAM_NAME, PACKAGE_NAME, 
        Version, AUTHORS, (char *) NULL);
    }
  return EXIT_STATUS;
}
\end{lstlisting}
\begin{lstlisting}[caption={An excerpt of the \texttt{true} 1999, with no variability}, label=truefalse02, language=C]
int main ()
{
  exit (0);
}
\end{lstlisting}

This is the case only for their run-time variability. But, as their size in LoC has changed among the versions, we decided to investigate in depth the source code of the two smallest programs, \texttt{true} and \texttt{false}, to understand the nature of these size changes and check for any differences in their compile-time options.
In one of the earliest \coreutils releases, version 5.0 (2003)\footnote{The source of analysed release: \url{https://mirror.cyberbits.eu/gnu/coreutils/}}, \texttt{true} had 43~LoC, whereas \texttt{false} did not yet have a source file. 
\Cref{truefalse} shows an excerpt of \texttt{true}'s implementation. 
In version~5.0, it had only line~6 from lines~5-9. The program name \texttt{"true"}, which will be bound at compile-time, but its value remained the same, that is, it did not vary.
From version 5.90 onward, when \texttt{false} had its own source file for the first time, developers added a few more LoC to \texttt{true}, among which also lines 5 and 7-9.   
Thus, in addition to the existing run-time variability (\cf lines~13-20), some compile-time variability was added. 
Currently, the latest versions of \texttt{true} and \texttt{false} have twice as much variability, with 2 compile-time options and 2 run-time options each.
Out of curiosity, we went further back in time and noticed that the implementation of \texttt{true} in 1999 has had only 4 LoC and no variability, that is, neither run-time nor compile-time options. ~\Cref{truefalse02} shows its entire implementation.

\begin{lstlisting}[caption={\texttt{true} and \texttt{false} in Unix, 1979 (from: ~\cite{inflation2})}, label=truefalse03, language=C]
$ ls -l /bin/true /bin/false
<@\textcolor{codegreen}{>> -rwxr-xr-x 1 root root 0 Jan 10 1979 /bin/true}@>
<@\textcolor{codegreen}{>> -rwxr-xr-x 1 root root 7 Jan 10 1979 /bin/false}@>
\end{lstlisting}

\Cref{truefalse03} shows the respective binary size of \texttt{true} and \texttt{false} in their very first version (adopted from~\cite{inflation2}). Basically, \texttt{true} was an \textit{empty file}. The way how it worked is that the shell would open the empty file, do nothing, and exit with a true status code 0.
Whereas, \texttt{false} was a file with 7 characters, that is, `\texttt{exit 1 }` (including the line-feed at the end). It would return 1, which signifies failure~\cite{bintrue}. 
Thus, the initial versions of \texttt{true} and \texttt{false} had little to no lines of code (LoC), no variability, and yet retained the same exact functionality as their current version~(9.1).

These results show that while the run-time variability of most programs remained unchanged across releases, compile-time variability evolved notably, particularly in the smallest programs like \texttt{true} and \texttt{false}. Interestingly, these programs began with little to no variability (\aka null variability), yet over time, both compile-time options and lines of code increased, reflecting the incorporation of broader configurability for the same exact functionality. While \hyperref[RQ2]{$RQ_2$} highlighted a strong relationship between run-time variability and program size when analyzing their variability in space, the analyzed variability in time (\ie \hyperref[RQ3]{$RQ_3$}) underscores a relationship between compile-time variability and size, as observed through the evolution of two small programs. Extending this analysis to other subject programs could uncover broader trends, but this requires developing tools to automate compile-time variability identification in all program releases, which we leave for future work.

\begin{tcolorbox}[boxsep=-2pt]
\textbf{\hyperref[RQ3]{$\boldsymbol{RQ_{3}}$} insights:}
Early versions of some small-scale software exhibited minimal variability, often with no configuration options. Over time, variability increased, mostly due to the addition of compile-time options. Moreover, while run-time variability was correlated to program size in space, compile-time variability seems to be linked with size in time. These results highlight the importance of further exploring compile-time options and their impact on software growth over time.
\end{tcolorbox}

\subsection{Used Approached for Reducing Variability}
To answer our fourth research question (\hyperref[RQ4]{$RQ_4$}), we analyzed \toybox~\cite{toybox}, an alternative and lightweight implementation of \coreutils.

The \toybox's main objective is to provide the most common Linux programs into a single executable, which would be simple, small, and fast to turn resource constraint devices, such as Android, into a development environment~\cite{toybox}. 
To answer \hyperref[RQ4]{$RQ_4$}, we explored how much smaller the \toybox programs are compared to \coreutils and what developers did to achieve this reduction. While these programs are expected to be smaller in size, the question is: what happens to their variability? Is it preserved or reduced?

To begin with, the recent release of \toybox\footnote{The analysed release of \toybox: \url{https://github.com/landley/toybox/tree/0.8.8}} includes 227 programs and has 72\,300 LoC, or $90,53\%$ less than \coreutils.
Then, 85 of 108 programs in \coreutils are also in \toybox. 
Indeed, comparing their individual size showed that each of them have less LoC and smaller executable than their equivalent program in \coreutils. 
Due to space constraints, detailed data are provided in the shared repository. Regarding their run-time variability, we noticed that:
\begin{itemize}
    \item 6 programs have one or two more options than in \coreutils.
    \item 19 programs have the same number of options.
    \item 60 programs exhibit less variability in \toybox.
\end{itemize}
Based only on run-time variability, these data suggest that programs in \toybox have become smaller by mainly reducing their variability ($71\%$ of them). But, the question is how did they achieve that goal?

\begin{table}[t]
\caption{\texttt{cat}'s size and variability in \coreutils and \toybox}
\label{tab:cat}
\begin{tabular}{@{}lrrrr@{}}
\toprule
\textbf{\texttt{cat}} & LoC & Size & Compile-time & Run-time \\ \midrule
\coreutils & 526 & 219,02 [KiB] & 4 & 12\\
\toybox & 40 & 18,14 [KiB] & 0 & 4 + 2 \\ \bottomrule
\end{tabular}
\end{table}

\subsubsection{Removing Variability}
In order to get a glimpse of understanding which variability is reduced and how, we chose to compare the implementation of \texttt{cat} program in \coreutils and \toybox.
It is one of the most used programs\footnote{Based on UNIX user data set: \url{http://archive.ics.uci.edu/ml/datasets/unix+user+data}}, which copies (concatenates) files to standard output. \Cref{tab:cat} shows that its implementation in \toybox has $92.40\%$ less LoC,  a $91.72\%$ smaller executable, and $62.50\%$ less compile-time and run-time options (\ie variability). 

Basically, there is no visible compile-time variability in the implementation of \texttt{cat} in \toybox, that is, the built \texttt{cat} in \toybox is always the same or is of null variability at compile-time.
Then,~\Cref{catcoreutils,cattoybox} show that \texttt{cat} has also 6 less run-time options in \toybox than in \coreutils. 
As it can be noticed, this is achieved by removing the unnecessary variability. 
First, the long options are removed, that is, only their abbreviations are kept. For example, \mbox{-}\mbox{-}\texttt{show}\mbox{-}\texttt{nonprinting} is removed and \mbox{-}\texttt{v} is kept (\cf lines 14-15 in~\Cref{catcoreutils} and lines 6-8 in~\Cref{cattoybox}). Secondly, duplicated options are removed, such as \mbox{-}\texttt{b} and \mbox{-}\texttt{n} (\cf lines 4-5, 8 in~\Cref{catcoreutils}). Plus, no  \mbox{-}\texttt{s}  option exists in \toybox. It is the only one that is completely removed.
Thirdly, the equivalent options are removed too, namely \mbox{-}\texttt{A}, \mbox{-}\texttt{e}, and \mbox{-}\texttt{t} (\cf lines 3, 6, 11 in~\Cref{catcoreutils}). 
They are equivalent with the combination of other options, hence they are completely removed in \toybox.
Lastly, although the \mbox{-}\mbox{-}\texttt{help} and \mbox{-}\mbox{-}\texttt{version} are understood by \texttt{cat} in \toybox (also by most of its other programs), they are invisible here but contribute to its variability (\cf~\Cref{tab:cat}).   

\begin{lstlisting}[style=ontop,escapechar=\%, caption={The run-time option of \texttt{cat} in \coreutils 9.1}, label=catcoreutils]
%\HilightE%[coreutils9.1]$ ./src/cat --help
>> Usage: ./src/cat [OPTION]... [FILE]...
%\HilightD%>> -A, --show-all          equivalent to -vET
%\HilightD%>> -b, --number-nonblank   number nonempty output lines, 
%\HilightD%>>                         overrides -n
%\HilightD%>> -e                      equivalent to -vE
%\Hilight%>> -E, --show-ends         display $ at end of each line
%\HilightD%>> -n, --number            number all output lines
%\HilightB%>> -s, --squeeze-blank     suppress repeated empty output 
%\HilightB%>>                         lines
%\HilightD%>> -t                      equivalent to -vT
%\Hilight%>> -T, --show-tabs         display TAB characters as ^I
%\Hilight%>> -u                      (ignored)
%\Hilight%>> -v, --show-nonprinting  use ^ and M- notation, except 
%\Hilight%>>                         for LFD and TAB
%\HilightC%>>     --help        display this help and exit
%\HilightC%>>     --version     output version information and exit
\end{lstlisting}
\begin{lstlisting}[style=ontop,escapechar=\%, caption={The run-time option of \texttt{cat} in \toybox 0.8.8}, label=cattoybox]
%\HilightE%[toybox]$ ./install/bin/cat --help
>> usage: cat [-etuv] [FILE...]
%\Hilight%>> -e      Mark each newline with $
%\Hilight%>> -t      Show tabs as ^I
%\Hilight%>> -u      Copy one byte at a time (slow)
%\Hilight%>> -v      Display nonprinting characters as escape 
%\Hilight%>>         sequences with M-x for high ascii characters 
%\Hilight%>>         (>127), and ^x for other nonprinting chars
\end{lstlisting}

In~\Cref{catcoreutils,cattoybox} also can be visually distinguished the implemented (\tikz\draw[black,fill=codegreen] (0,0) rectangle (0.2,0.2);), removed (\tikz\draw[black,fill=red!40] (0,0) rectangle (0.2,0.2);), unimplemented (\tikz\draw[black,fill=yellow!70] (0,0) rectangle (0.2,0.2);), and invisible (\tikz\draw[black,fill=codepurple!30] (0,0) rectangle (0.2,0.2);) options in \toybox.
Although other and different reasons for keeping or removing some variability may exist, the illustrative example of \texttt{cat} shows one way to reducing, and in some cases nullifying, the variability of a software program. By \textit{nullifying} we mean that the program's variability is reduced, sometimes entirely, while still delivering the same essencial value to users. Additionally, we identified various examples of programs in \toybox where variability has been completely removed (\ie nullified), such as \texttt{sync}, \texttt{true}, \texttt{false}, and \texttt{sleep} (though \texttt{sleep} still accepts different times as inputs).

\begin{minipage}[t]{3.7cm}
\begin{lstlisting}[caption={\texttt{true} in \toybox}, label=truetoybox, language=C]
#include "toys.h"
void true_main(void)
{
  return;
}
\end{lstlisting}
\end{minipage}
\hfill
\begin{minipage}[t]{4cm}
\begin{lstlisting}[caption={\texttt{false} in \toybox}, label=falsetoybox, language=C]
#include "toys.h"
void false_main(void)
{
  toys.exitval = 1;
}
\end{lstlisting}
\end{minipage}

\subsubsection{Resolving Variability}
Looking into the implementation of \toybox, we also noticed that some of its programs have less variability because it is resolved earlier. Concretely, \texttt{true} and \texttt{false} programs in \toybox have only 5 LoC each and no compile-time or run-time variability.
Semantically, these programs are opposites, which may explain why their implementation in \coreutils has undergone several changes, as discussed in~\Cref{variabilityovertime}. \Cref{truetoybox,falsetoybox} show their current and respective implementation in \toybox. By comparing the two implementations of \texttt{true} it can be noticed that the compile-time variability in lines 5-9 in~\Cref{truefalse} is resolved in~\Cref{truetoybox,falsetoybox}. As a result of this, now each of two programs are into their own source files, \texttt{true.c} and \texttt{false.c}, without the implemented logic by preprocessor directives to build them from a common implementation. 
Despite this, both programs in \toybox together have less LoC and binary size than in \coreutils. 

Moreover, we observed that while run-time variability in \toybox is reduced, and in some cases nullified by removing unnecessary and duplicate options, compile-time variability is primarily reduced, or even nullified, by resolving it during programming-time.

\begin{tcolorbox}[boxsep=-2pt]
\textbf{\hyperref[RQ4]{$\boldsymbol{RQ_4}$} insights:}
Based on small-scale software, software systems can become even smaller or lightweight by \emph{removing} redundant and unnecessary run-time variability, consolidating functionalities, and \emph{resolving} compile-time variability early. 
\end{tcolorbox}

\section{Discussion}
\label{discussion}
These results highlight the extensive configuration space in small-scale software, despite their size, and the strong correlation between their run-time variability and size. Over time, variability in these software seems to grow mostly due to the added compile-time variability. The observed results suggest that changing certain options from run-time to compile-time could lead to notable reductions in binary size or LoC. In fact, the broader impact of altering the binding time of configuration options on binary size seems to be an important area for further investigation. Such a study could help developers to achieve a better balance between variability and software size, while also identifying and eliminating unused features that contribute to unnecessary bloat. Moreover, this strategy could pave the way for creating more specialized software systems, as opposed to those bloated with redundant or rarely used features.

It is important to note that we do not rule out the possibility that additional methods for reducing variability could exist. For instance, the disparity in the number of programs between \toybox (227) and \coreutils (108) is important to investigate. It raises the question of whether some programs in \coreutils have been split into multiple, more specialized programs in \toybox. Still, the overall codebase of \toybox remains significantly smaller than that of \coreutils. This observation suggests potential differences in the underlying used algorithms or programming methodologies, with \coreutils potentially favoring a more verbose approach. 

As opposed to the SPL variants that might be used independently of one another, the \coreutils programs are frequently used together, to perform sequential tasks, in scripting and command-line workflows, such as \texttt{ls} and \texttt{cd}. Identifying common combinations of these programs and evaluating the variability contributed by each could help to simplify workflows by removing unnecessary options in some typical use cases. Moreover, we believe that similar findings based on small-scale software offer valuable insights that could be applied to larger software with more extensive codebases.

\section{Unveiling Null Variability}
\label{definitionnull}
Building on the findings from our research questions, we introduce \emph{null-variable} software systems, providing a definition and outlining five observed properties of \emph{null variability}. We deliberately use the term "null" instead of "zero" or "no" variability since a configurable software is often described by the \textit{set} of features or configuration options 
that it contains. Hence, we use \textit{null} for variability similar to the \textit{null set} (\ie $\{\}$) in the set theory. 

\subsection{Notations}
To begin, we present the necessary notations to formally define our concept.
Let $S$ be a set of all existing software systems and $s \in S$ be a given software system.
If $s$ is a software system with variability, then let $F_s$ be the set of all its features or configuration options. 
Each of those options can be on/off or have multiple values. 
When combined, the set of options used by a user constitutes \textit{a system configuration}, which yields a unique instance (variant) of the system. Furthermore, an option's value is set at a specific point in time, known as its binding time. Therefore, let $T =  \{..., t_p, t_c, t_r, ...\}$ be the set of possible binding times of the option $f \in F_s$, where $t_p$ is the binding at programming-time, $t_c$ is the binding at compile-time, and $t_r$ is the binding at run-time.  
In this work, we consider that all mandatory features are bound or resolved before compile-time, \ie at $t_p$. 
Conversely, varying features (\ie~configuration options) are resolved at one of two subsequent binding times, namely, $t_c$ or $t_r$.

\subsection{Properties}
\label{properties}
Based on the obtained results in this study, we derived the following five properties of software systems of null variability.

\paragraph{\textbf{$P_1$: Existing software}}
A software system of null variability, denoted as $s_{null}$, is an existing, complete, and functional piece of software. It is different from no software at all. Then, just like all other kinds of working or complete software, an unvarying program (\ie of null variability) delivers values to the end-users. This means that it is not a running but featureless software, in the sense that it does nothing.  This may appear to be a paradoxical statement, but all features in a null-variable software are 
mandatory, in the sense that they cannot be selectively enabled or disabled after implementation. As $F_s$ is the set of all features in a software system $s$, then formally

\begin{equation} \label{eq1}
\begin{aligned}
s_{null} \in S \\
F_{s_{null}} \neq \emptyset \text{ , }
F_{s_{null}} = \{f_x \mid bound(f_x) \le t_p, x \in \mathbb{N}\} \\
\end{aligned}
\end{equation}

\paragraph{\textbf{$P_2$: Multiplicity across domains}}
Null variability is not limited to a single software system or domain. Multiple systems can be of null variability, regardless of the programming paradigm or application domain. Formally,

\begin{equation} \label{eq2}
\begin{aligned}
    s_{null} \in S_{null} \text{ , where } S_{null} \subset S \text{ and } 	\lvert S_{null} \rvert \ge 0 
\end{aligned}
\end{equation}

\paragraph{\textbf{$P_3$: Behavioral consistency}} 
A null-variable software system behaves as a constant system. For any input $I$ it produces a consistent, though not necessarily identical, output $U$. This property ensures predictable, unchanging behavior across all contexts. Formally,

\begin{equation} \label{eq3}
\begin{aligned}
    s_{null}(I) = U
\end{aligned}
\end{equation}

\paragraph{\textbf{$P_4$: Distinction from the core part in an SPL}}
In variability-intensive software systems or SPLs, the core (or base) part is often distinguished from the commonalities ($C_s$) and variabilities ($V_s$).  While the core part is often understood as the constant part among the possible software products that can be derived from a codebase, it may still include some variability. Additionally, core features may represent a basis (\eg classes or interfaces) that is extended later and, therefore, not yet a complete, running software. In contrast, a null-variabile software ($s_{null}$) is a complete, unvarying software where all features ($F_{s_{null}}$) are compulsory and bound before or at programming time ($t_p$). This distinction is formalized as:

\begin{equation} \label{eq4}
\begin{aligned}
    F_s = C_s \cup V_s \text{ , } C_s \cap V_s = \emptyset\\
    \forall c \in C_s \text{ , } bound(c) \leq t_p \text{ and }
    \forall v \in V_s \text{ , } bound(v) \in \{t_c, t_r\} \\
    F_{s_{null}} = \{f_x \mid bound(f_x) \le t_p, x \in \mathbb{N}\} \\
    F_{s_{null}} \neq \emptyset \text{ , } C_{s} \subset F_{s_{null}} \text{ , where } C_{s} \neq \emptyset \\
\end{aligned}
\end{equation}

\paragraph{\textbf{$P_5$: Binding time constraints}}
The concept of null variability can be used also to denote the absence of variability at a particular binding time. For instance, a software with no run-time configuration options is \textit{null-varying at run-time}. Similarly, a \textit{null-varying software at compile-time} has no variability in compile time. 

\subsection{Definition}
Hence, we define \emph{null software variability} as the exact opposite of software variability, as described by \citet{capilla2013systems}. 

\begin{definition} \label{def2}
Null variability in software is the deliberate absence of the ability to customize or configure the software at or after a certain binding time. A software with null variability delivers consistent and unchanging behavior regardless of its context or environment.
Formally, if $t_b$ is the fixed binding time, then
\begin{equation} \label{eq5}
s_{null} \Leftrightarrow F_s \neq \emptyset \text{ and } \forall f \in F_s \text{ , } bound(f) \leq t_b
\end{equation}
\end{definition}

\section{Threats to validity}
\label{threats}
\paragraph{Internal threats}
A first internal threat stems from the identification of compile-time and run-time options of programs. Their variability can be different in case that somebody else uses a slightly different definition of what an options is, compared to our given explanation in~\Cref{protocol}. As for the compile-time options, we plan to use the state-of-the-art approaches for identifying the implemented variability by preprocessors in C (\eg~\cite{10.1145/3729423}). Still, as we studied in depth most of the programs, we manually crosschecked their amount of variability.
In addition, other variation points may exist in code, which may have other binding times. Moreover, we do not consider the variability of external libraries used by a program or in build scripts. But, this does not present a direct threat to our study. 
Another threat is related with the counting of LoC, because of the SPL nature of \coreutils and \toybox. To increase its precision we used the complementary and alternative metric of binary size. Importantly, in both cases we got the same results in our research questions. To mitigate these threats, all the data are extracted by two authors and crosschecked by two other authors.

\paragraph{External threats}

While we believe that analysing 108 programs of \coreutils, and 85 programs of \toybox that intersect with them, are strong evidence for our obtained results, still we considered only the C-based and small programs. 
While our study showed that the empty file version of the \texttt{true} program is the smallest null-variable software program we encountered, it is important to note that we cannot determine the upper limit of the size of a null-variable software program based solely on these findings. Still, whatever the size of a software system, we believe that the concept of null variability remains valid for them too, and also similar approaches can be employed to reduce their variability.

\section{Related work}
\label{related}
Most research on software variability focuses on large-scale systems, such as the Linux kernel~\cite{10.1145/3307630.3342400,10.1145/3729423}. Then, many studies also highlight the increasing variability in software systems~\citep{xu2015hey,alenezi2015empirical,10.1145/1966445.1966451}, often adding complexity through numerous configuration options. In contrast, our work examines the variability of small-sized software systems, that is, those with few lines of code or small executables. These systems are intentionally kept simple yet effective, delivering real, everyday value to users. Still, to the best of our knowledge, the extent of their variability and how it is linked with their size has not been explicitly explored in prior research.

However, numerous subdomains related to SPL engineering are linked to this work. For instance, \citeauthor{el2019metrics}~\cite{el2019metrics} conducted a systematic literature review and presented a catalog of over 140 variability-aware implementation metrics for software product lines, including "$\#$ of configuration options" and "$\#$ of  features" in an SPL. In this work, we quantified the variability of 108 small to tiny programs using existing metrics and introcude a new metric, that of \textit{null variability at compile-time or run-time}. While \citeauthor{el2019metrics} focused on identifying relevant variability metrics in the literature, our objective was to apply these metrics to analyze and characterize the variability of real-world, small-sized programs, including how their variability evolves over time.

There are different works that explore how the variability in C-based systems is addressed, managed, or extracted. Our work examines how much variability is included in small-sized systems, showing that it can still be significant.  We also highlight that, despite their size, there are real scenarios where this variability needs further reduction, and introduce the notion and characteristics of systems with null variability.
On the other hand,~\citeauthor{10.1145/3168365.3168371}~\cite{10.1145/3168365.3168371} conducted a study that, while not directly related, parallels ours in exploring an unconventional aspect of software systems. They focused on the commonality of software systems instead of variability, whereas we study the variability of small-size software systems.

Furthermore, several other topics in software engineering may appear related, including the \textit{low code/no code}~\citep{rokis2022challenges} or the KISS~\citep{vidal2011yahsp2} philosophies, which emphasize keeping the codebase as small and simple as possible during development. Other related areas include \textit{software debloating}~\citep{10.1145/3379469,ternava2022specialization}, which involves removing superfluous features, options, or libraries at the code level to reduce the binary size and surface attack of systems; \textit{optimisation} or \textit{restructuration}~\citep{4339264,4145034}, aimed at removing rarely or never used features to decrease the complexity of systems; and \textit{dependency reduction}~\citep{sinnema2004covamof}, which focuses on minimizing unnecessary dependencies.

Some less related works, but close to the protocol of our experiments, are those on the growing complexity of software systems~\citep{alenezi2015empirical}, extracting compile-time variability out of preprocessor directives~\citep{10.1145/2364412.2364428,10.1145/2076021.2048128,10.1145/3729423}, and 
analysing the evolution of variability~\citep{10.1145/2491627.2491645}.

\section{Conclusion}
\label{conclusion}
While the extent, implementation, management, and testing of software variability are well-recognized challenges in modern (ultra) large-scale software systems, little attention has been given by researchers to the extent of variability in small-scale software.
We conducted an empirical investigation into the extent of variability in small-scale software and introduced the concept of null variability.
By analyzing 108 programs from GNU \coreutils, we revealed that even small-scale software, averaging $308$ [KiB], exhibits notable configurability, with an average configuration space of $2^{15}$ and offering up to 76 options per program. This shows that despite their modest size, such software presents similar challenges in testing and maintenance as large-scale software systems.
Our findings also showed a high correlation ($0.78$) between run-time variability and software size. Then, an analysis of 84 releases of 20 \coreutils programs showed that early versions of programs exhibited little to no variability, while a focused study in smallest programs highlighted that compile-time variability have an effect in the codebase size of small-scale software over time. These results highlight how design decisions and the binding time of variability influence the codebase size both at specific points in time and over its development. 
Additionally, we show that software can be made more lightweight by removing redundant run-time variability or resolving some of the compile-time variability early in the development process.

Finally, we introduced \textit{null variability} as a formal concept to describe software systems with no configurability beyond mandatory features. This concept provides a new way on describing and designing immutable, minimal, specialized, or lightweight software systems. We hope these insights encourage further exploration of variability reduction in both small and large-scale systems, for improving software maintainability and usability.

\balance 
\bibliographystyle{ACM-Reference-Format}
\bibliography{acmart}

\end{document}